\documentclass[aps,prd,preprintnumbers,amsmath,amssymb,nofootinbib,11pt]{revtex4}
\usepackage{eepic}
\usepackage{indentfirst}
\usepackage{mathrsfs}
\usepackage{fancyhdr}
\usepackage{ulem}
\usepackage{float}
\usepackage{graphicx}
\usepackage[
colorlinks=true, linkcolor=black, breaklinks=true, urlcolor=blue,
citecolor=green]{hyperref}
\usepackage{epstopdf}
\usepackage{bm,bbm}
\usepackage{amsmath}
\usepackage{amssymb}

\usepackage{tabularx}
\usepackage{subfigure}
\usepackage{epsfig}
\usepackage{color}
\usepackage{slashed}
\usepackage{hyperref}

\newcommand{\be}{\begin{equation}}
\newcommand{\ee}{\end{equation}}
\newcommand{\bea}{\begin{eqnarray}}
\newcommand{\eea}{\end{eqnarray}}

\newcommand{\bra}{\langle}
\newcommand{\ket}{\rangle}

\newcommand{\dilog}{\mathrm{Li}_2}

\allowdisplaybreaks

\begin{document}
\pagestyle{plain}

\title {\boldmath\Large Momentum dependence of $\rho-\omega$ mixing in the pion vector form factor and its effect on $(g-2)_\mu$ }
\author{ Yun-Hua~Chen}\email{yhchen@ustb.edu.cn}
\author{ Meng-Ge Qin}
\affiliation{School of Mathematics and Physics, University
of Science and Technology Beijing, Beijing 100083, China  }

\begin{abstract}

The inclusion of the $\rho-\omega$ mixing effect is essential for a precise description of the pion electromagnetic form factor in the $e^+e^- \rightarrow
\pi^+\pi^-$ process, which quantifies the two-pion contribution to the anomalous magnetic moment of the muon $a_\mu$. 
In this paper, we analyse the momentum dependence of the $\rho-\omega$ mixing by considering loop contributions at the next-to-leading order in the $1/N_C$ expansion within the framework of resonance chiral theory. 
 We revisit the work {[}Y. H. Chen, D. L. Yao, and H. Q. Zheng, Commun. Theor. Phys. 69 (2018) 1{]}, considering the contribution arising from the kaon mass splitting in the kaon loops and the latest experimental data. We perform two kinds of fits (with momentum-independent or momentum-dependent $\rho-\omega$ mixing amplitude) to describe the $e^+e^-\rightarrow \pi^+\pi^-$ and $\tau\rightarrow \nu_{\tau}2\pi$ data within the energy region of 600$\sim$900 MeV and the decay width of $\omega \rightarrow \pi^+\pi^-$, and compare their results. Our findings indicate that both the momentum-independent and momentum-dependent $\rho-\omega$ mixing schemes provide appropriate descriptions of the data. However, the momentum-dependent scheme exhibits greater self-consistency, considering the reasonable imaginary part of the mixing matrix element $\Pi_{\rho\omega}$ obtained. Regarding the contribution to the anomalous magnetic moment of the muon, $a_\mu^{\pi\pi}|_{[0.6,0.9]\text{GeV}}$, the results obtained from the fits considering the momentum-dependent $\rho-\omega$ mixing amplitude agree well with those obtained without incorporating the momentum dependence of the $\rho-\omega$ mixing, within the margin of errors. Furthermore, based on the fitted values of the relevant parameters, we observe that the decay width of $\omega \rightarrow \pi^+\pi^-$ is predominantly influenced by the $\rho-\omega$ mixing effect.

\end{abstract}

\maketitle

\newpage

\section{Introduction}

The anomalous magnetic moment of the muon, denoted as $a_\mu=(g_\mu-2)/2$, plays an crucial role in the precision tests of the Standard Model (SM)~\cite{Aoyama:2020ynm,Colangelo:2022jxc}. The long-standing discrepancy between the SM prediction of $a_\mu$ and its experimental measurement recently has been updated to 4.2 standard deviations~\cite{Muong-2:2006rrc,Muong-2:2021ojo} and it has sparked numerous theoretical investigations.
The SM uncertainty on $a_\mu$ is dominated by hadronic vacuum polarization (HVP), with the largest contribution originating from the $\pi\pi$ intermediate states, accounting for over 70\% of the HVP contribution.
Theoretically, the two-pion low-energy contribution to $a_\mu$ is expressed as an integral over the modulus squared of the pion electromagnetic form factor, which can be extracted from the $e^+ e^--$annihilation experiments. In principle,
the two-pion contribution to $a_\mu$ can be evaluated accurately as long as the experimental data of $e^+ e^- \to \pi^+\pi^-$ are available everywhere at the required level of precision. While it is known that a tension exists between the two most precise measurements by BaBar and KLOE Collaborations: the BaBar data lie systematically above the KLOE results in the dominant $\rho$ region. Consequently, considerable efforts have been dedicated to finely describing the pion electromagnetic form factor~\cite{Colangelo:2018mtw,Colangelo:2020lcg,Colangelo:2022prz,Davier:2019can,Qin:2020udp,Yi:2021ccc,Ananthanarayan:2016mns}. In the dominant $\rho$ region of the $e^+ e^- \to \pi^+\pi^-$ process, the isospin-breaking effect due to $\rho-\omega$ mixing, which becomes enhanced by the small mass difference between the $\rho$ and $\omega$ mesons, plays a significant role and needs to be considered properly.

Usually, the momentum dependence of the $\rho-\omega$ mixing amplitude is neglected, and a constant mixing amplitude is used to describe the $e^+ e^- \to \pi^+\pi^-$
data due to the narrowness of the $\omega$ resonance. The first study of the momentum dependence of the $\rho-\omega$ mixing amplitude was conducted by Ref.~\cite{Goldman}. Based on a quark loop mechanism of the $\rho-\omega$ mixing, it was found that the mixing amplitude
significantly depends on momentum.
Subsequently, the investigation of various loop mechanisms for $\rho-\omega$ mixing was initiated in different models, such as the global color model
~\cite{Mitchell}, extended
Nambu-Jona-Lasinio (NJL) model~\cite{Shakin,Braghin:2020enw}, the chiral
constituent quark model~\cite{MLYan98,MLYan00}, and the hidden local symmetry
model~\cite{Benayoun00,Benayoun01,Benayoun08}. In our pervious work~\cite{Chen:2017jcw}, we studied the $\rho-\omega$ mixing in a model
independent way by invoking Resonance Chiral
Theory (R$\chi$T)~\cite{Ecker}.
Guided by chiral symmetry and large $N_C$ expansion, R$\chi$T provides us
a reliable theoretical framework to study the dynamics with both light flavor resonances and pseudo-Goldstone mesons in the intermediate energy region~\cite{Guo:2011pa,Jamins,Roig:2013baa,chen2012,chen2014,Chen:2014yta}, and it has been successfully applied in the calculation of $a_\mu$ in the
SM~\cite{Pich:2001pj,Cirigliano:2002pv,Kampf:2011ty,Roig:2014uja,Guevara:2018rhj,Roig:2019reh,Miranda:2020wdg,Qin:2020udp,Arroyo-Urena:2021nil}.
In Ref.~\cite{Chen:2017jcw}, we calculated the one-loop contributions to the $\rho-\omega$ mixing, which are at
the next-to-leading order (NLO) in the $1/N_C$ expansion~\cite{Hooft,Rosell04,Cirigliano:2003yq,Guo:2014yva,Roig:2013baa}. In this article, we update the previous work of Ref.~\cite{Chen:2017jcw} by incorporating the contribution arising from the kaon mass splitting in the kaon loops.

Moreover, we focus on analysing the impact of the momentum dependence of $\rho-\omega$ mixing on describing the pion vector form factor data and its contribution to $a_\mu$. Specifically, we perform two types of fits (with momentum-independent or momentum-dependent $\rho-\omega$ mixing amplitude) describing the $e^+e^-\rightarrow \pi^+\pi^-$
and and $\tau\rightarrow
\nu_{\tau}2\pi$ data in the energy
region of 600$\sim$900 MeV, the decay width of $\omega \rightarrow \pi^+\pi^-$, and compare their results.
Our fit results demonstrate that both the momentum-independent and momentum-dependent $\rho-\omega$ mixing schemes can effectively describe the data, while the momentum-dependent scheme exhibits greater self-consistency due to the reasonable imaginary part of the extracted mixing matrix element $\Pi_{\rho\omega}$. Regarding the contribution to the anomalous magnetic moment of the muon, $a_\mu^{\text{HVP,LO}}[\pi^+\pi^-]$, evaluated between 0.6 GeV and 0.9 GeV, the results obtained from fits considering the momentum-dependent $\rho-\omega$ mixing amplitude are in good agreement with those from fits that do not include the momentum dependence of $\rho-\omega$ mixing, within the margin of errors.

This paper is organized as follows. In Sec.~\ref{section.TheoreticalFramework}, we introduce the description of
$\rho-\omega$ mixing and elaborate on the calculation of
$\rho-\omega$ mixing amplitude up to
the next-to-leading order in the $1/N_C$ expansion.
In Sec.~\ref{section.Phenomenology}, the fit results are shown and
the related phenomenologies are discussed. A summary is given in
Sec.~\ref{section.Conclusions}.

\section{Calculations in resonance chiral theory}\label{section.TheoreticalFramework}

\begin{figure*}[ht]
\centering
\includegraphics[scale=0.6]{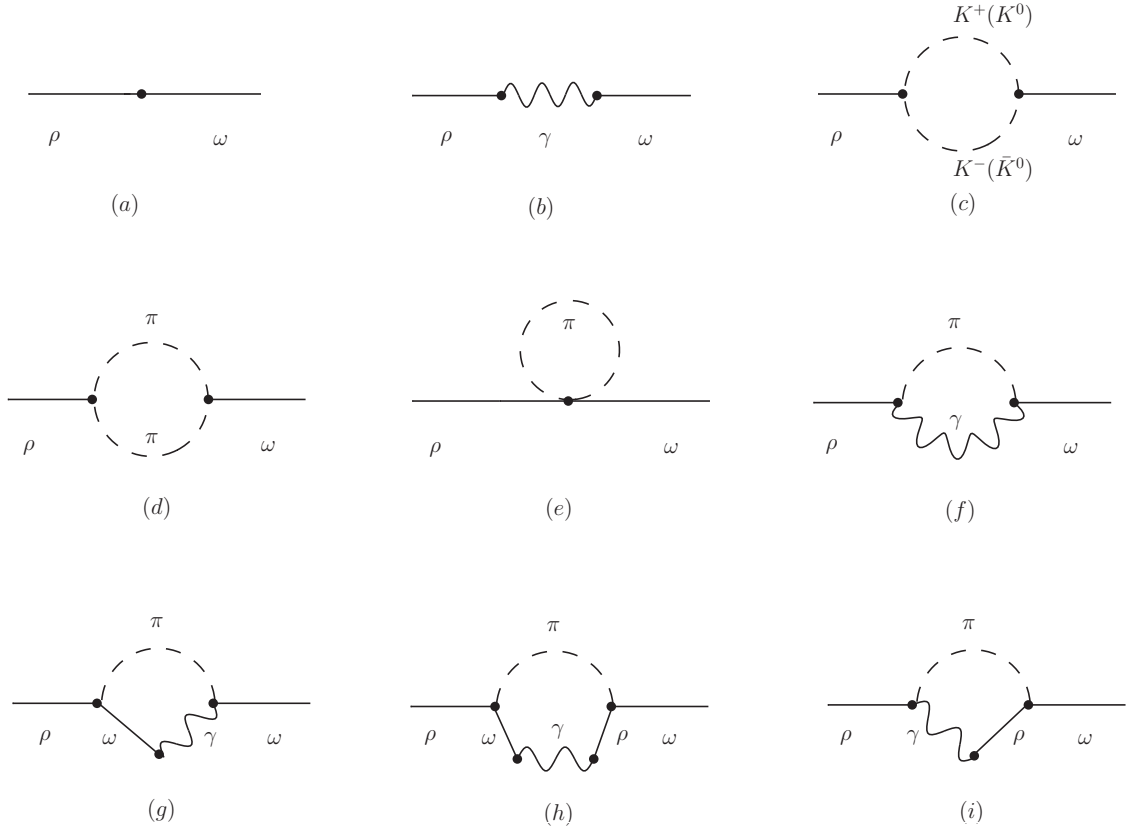}
\caption{ Feynman diagrams contributing to $\rho-\omega$
mixing. } \label{FeynmanDiagram}
\end{figure*}

In the isospin basis $|I,I_3\rangle$, we define the pure isospin states $|\rho_I\rangle\equiv |1,0\rangle$ and $|\omega_I\rangle\equiv |0,0\rangle$. The mixing between the isospin states of $|\rho_I\rangle$ and $|\omega_I\rangle$ can be implemented by considering the self-energy matrix
\bea \Pi_{\mu\nu}=T_{\mu\nu}\left(
                    \begin{array}{cc}
                      \Pi_{\rho\rho}(s) & \Pi_{\rho\omega}(s) \\
                      \Pi_{\rho\omega}(s) & \Pi_{\omega\omega}(s) \\
                    \end{array}
                  \right)\,,
\eea with $T_{\mu\nu}\equiv g_{\mu\nu}- \frac{p^\mu p^\nu}{p^2}$
 and $s\equiv p^2$.
The none-zero off-diagonal matrix element $\Pi_{\rho\omega}(s)$ contains the
information of $\rho-\omega$ mixing.
The mixing between the physical states of $\rho^0$ and $\omega$, is obtainable by introducing the following relation
\bea
  \left( \begin{array}{c}
          \rho^0 \\
          \omega
         \end{array} \right) =C \left( \begin{array}{c}
          \rho_I \\
          \omega_I
         \end{array} \right)\,,\qquad C= \left(\begin{array}{cc} 1 & -\epsilon_1 \\
\epsilon_2 & 1 \end{array} \right),\eea
where $\epsilon_1$ and $\epsilon_2$ are the mixing parameters. The matrix of dressed propagators corresponding to physical states is diagonal~\cite{Connell97},
         \bea
  \left( \begin{array}{cc}
          1/s_{\rho} & 0 \\
          0 & 1/s_{\omega}
         \end{array} \right)=C
  \left( \begin{array}{cc}
          1/s_{\rho} & \Pi_{\rho\omega}/s_{\rho}s_{\omega} \\
          \Pi_{\rho\omega}/s_{\rho}s_{\omega} & 1/s_{\omega}
         \end{array} \right)
 C^{-1}, \eea
where the abbreviations $s_\rho$ and $s_\omega$ are defined by the following:
 \bea\label{eq:srso} s_\rho&\equiv& s-\Pi_{\rho\rho}(s)-m_\rho^2,
\nonumber\\
s_\omega&\equiv& s-\Pi_{\omega\omega}(s)-m_\omega^2.
\eea
The information of $\rho-\omega$ mixing is encoded in
the off-diagonal element of the self-energy matrix, decomposed as follows:
\bea
\Pi_{\rho\omega}(s)=\Delta_{ud}\,S_{\rho\omega}(s)+4\pi\alpha\,E_{\rho\omega}(s)\
, \eea where $\Delta_{ud}=m_u-m_d$ is the mass difference between
$u$ and $d$ quarks, and $\alpha$ is the fine-structure constant. $S_{\rho\omega}(s)$ and $E_{\rho\omega}(s)$ represent the
structure functions of the strong and electromagnetic interactions,
respectively.  In this work, the diagrams in
Fig.~\ref{FeynmanDiagram} are calculated in R$\chi$T
up to NLO in $1/N_C$ expansion.

In R$\chi$T, the vector resonances can be described in terms
of antisymmetric tensor fields with the normalization
\begin{eqnarray}
\langle 0|V_{\mu\nu}|V,p\rangle =iM_V^{-1}\{p_\mu\epsilon_\nu(p)-p_\nu\epsilon_\mu(p)\}\,
\label{eqnormalization}
\end{eqnarray}
where $\epsilon_\mu$ denotes
the polarization vector.
Here the vector mesons are collected in a $3\times3$ matrix
\begin{equation}\label{defu3v}
V_{\mu\nu}=
 \left( {\begin{array}{*{3}c}
   \frac{1}{\sqrt{2}}\rho^0 +\frac{1}{\sqrt{2}}\omega & {\rho^+ } & {K^{\ast+} }  \\
   {\rho^- } & -\frac{1}{\sqrt{2}}\rho ^0 +\frac{1}{\sqrt{2}}\omega & {K^{\ast0} }  \\
   { K^{\ast-}} & {\overline{K}^{\ast0} } & -\phi   \\
\end{array}} \right)_{\mu\nu}\,.
\end{equation}

The effective Lagrangian for the leading order strong isospin-breaking effect, corresponding to the tree-level contribution diagram (a) in Fig.~\ref{FeynmanDiagram},
reads~\cite{Urech,Cirigliano:2006hb}
\begin{eqnarray}\label{eq.v8}
\mathcal{L}_2^{\rho\omega}=\lambda_6^{VV}\langle
V_{\mu\nu}V^{\mu\nu}\chi_+\rangle  \,,
\end{eqnarray}
with $\chi_+=u^{+}\chi u^{+}+ u\chi^{+}u$ and $\chi=2B_0(s+ip)$.
The pseudo-Goldstone bosons originating from the spontaneous breaking of chiral symmetry, can
be filled nonlinearly into
\begin{equation}
u_\mu = i \left( u^\dagger\partial_\mu u\, -\, u \partial_\mu
u^\dagger\right) \,, \qquad
u = \exp \Big( \frac{i\Phi}{\sqrt{2}F} \Big)\,,
\end{equation}
with the Goldstone fields
\begin{align}
\Phi &=
 \begin{pmatrix}
   {\frac{1}{\sqrt{2}}\pi ^0 +\frac{1}{\sqrt{6}}\eta  } & {\pi^+ } & {K^+ }  \\
   {\pi^- } & {-\frac{1}{\sqrt{2}}\pi ^0 +\frac{1}{\sqrt{6}}\eta} & {K^0 }  \\
   { K^-} & {\bar{K}^0 } & {-\frac{2}{\sqrt{6}}\eta}  \\
 \end{pmatrix} . \label{eq:u-phi-def}
\end{align}
Here $F$ is the pion decay constant.
Considering the mass relations of the vector mesons at
$O(p^2)$ in terms of the quark counting rule, the value of the coupling constant is determined: $\lambda_6^{VV}=1/8$~\cite{Urech,Cirigliano:2006hb}.
Thus the tree-level strong contribution turns out to be
\begin{eqnarray}\label{eq.TreeAmplitude}
S_{\rho\omega}^{(a)}=2\,M_V \ .
\end{eqnarray}

The Lagrangian describing the interactions between $V_{\mu\nu}$ and electromagnetic fields or Goldstone bosons are given by
\begin{eqnarray}\label{eq.interactionRchiT}
\mathcal{L}_{2}(V)&=&\frac{F_V}{2\sqrt{2}}\langle V_{\mu\nu}f_{+}^{\mu\nu}\rangle+\frac{iG_V}{\sqrt{2}}\langle V_{\mu\nu}u^\mu
u^\nu\rangle\ ,
\end{eqnarray}
with the relevant building blocks defined by
\begin{eqnarray}
 u_\mu &=&i[u^{+}(\partial _\mu \
-ir_\mu)u-u(\partial _\mu -il_\mu)u^{+}]\,, \nonumber\\
f_{\pm}^{\mu\nu}&=&uF_L^{\mu\nu}u^{+}\pm u^{+}F_R^{\mu\nu}u \,.
\end{eqnarray}
Here $F_{L,R}^{\mu\nu}$ are field strength tensors composed of the left and
right external sources $l_\mu$ and $r_\mu$, and $F_V$ and $G_V$ are real
resonance couplings constants.
The tree-level electromagnetic contribution from diagram (b) in
Fig.~\ref{FeynmanDiagram} can be calculated by
using the Lagrangian in Eq.~\eqref{eq.interactionRchiT}:
\begin{eqnarray}\label{eq.TreeAmplitude}
E_{\rho\omega}^{(b)}=\frac{ F_\rho F_\omega}{3}\,.
\end{eqnarray}
The physical decay constants $F_\rho$ and $F_\omega$ have been employed in the amplitude, and are
differentiated by means of isospin breaking.

The loop contributions of diagrams (d)-(i) in Fig.~\ref{FeynmanDiagram} have been extensively discussed in our previous work~\cite{Chen:2017jcw}. However, a noteworthy distinction in our current study is the inclusion of the contribution from diagram (c), which arises from the kaon mass splitting within the kaon loops. To ensure comprehensiveness, we present the expressions for the loop contributions in the Appendix~\ref{appendix-loop-contribution}.
Furthermore, it should be noted that the ultimate expression for the renormalised mixing amplitude ${\Pi}_{\rho\omega}(p^2)$ is presented in Equation (A23).

\section{Phenomenological discussion} \label{section.Phenomenology}

The mass and width of the $\rho$ meson are conventionally determined by fitting to the experimental data of $e^+e^-\rightarrow \pi^+\pi^-$ and
$\tau\rightarrow \nu_{\tau}2\pi$~\cite{ParticleDataGroup:2020ssz}, where various mechanisms are used to describe the $\rho-\omega$ mixing effect. To prevent interference caused by their $\rho-\omega$ mixing mechanisms, we treat the mass $M_\rho$, the relevant
couplings $G_\rho$ and $F_\rho$ as free parameters in our fit.
Regarding its width, the energy-dependent form is constructed in the way introduced in~\cite{Dumm}:
\begin{eqnarray}
\Gamma_{\rho}(s)=\frac{sM_\rho}{96\pi
F^2}\left[\sigma_\pi^3\theta(s-4m_\pi^2)+\frac{1}{2}\sigma_K^3\theta(s-4m_K^2)\right]\,,
\end{eqnarray}
where $\sigma_{P}  \equiv  \sqrt{1-4m_{P}^{2}/s}$, and $\theta(s)$ is the step function.

For the $\omega$ mass, it has been pointed out in Refs.~\cite{Colangelo:2018mtw,Colangelo:2022prz} that the result determined from $e^+e^- \rightarrow
\pi^+\pi^-$ is inconsistent with that from particle data group (PDG)
~\cite{ParticleDataGroup:2020ssz}, primarily determined by experiments involving $e^+e^- \rightarrow 3\pi$ and $e^+e^- \rightarrow \pi^0\gamma$.
Therefore, we treated the $\omega$ mass and width as free parameters and estimated them by fitting in our programme.
The physical coupling $F_\omega$ can be determined from the
decay width of $\omega\rightarrow e^+e^-$. Using the Lagrangian formula
in Eq.~\eqref{eq.interactionRchiT}, one can derive the decay width
\begin{eqnarray}
\Gamma_{\omega}^{e^+e^-}=\frac{4\alpha^2\pi
F_\omega^2(2m_e^2+M_\omega^2)\sqrt{M_\omega^2-4m_e^2}}{27
M_\omega^4}\,,
\end{eqnarray}
and obtain the expression of $F_\omega$.
With the decay widths given above, $s_\rho$ and $s_\omega$ in Eq.~\eqref{eq:srso} can be rewritten as
\bea
s_\rho&\simeq& s-M_{\rho}^2+iM_{\rho}\Gamma_{\rho}(s) \ ,\nonumber\\
s_\omega&\simeq& s-M_{\omega}^2+iM_{\omega}\Gamma_{\omega}\ .
\eea

The pion form-factor in $\tau\rightarrow \nu_{\tau}2\pi$ decay, irrelevant to $\rho-\omega$ mixing effect, were thoroughly
studied in Refs.~\citep{Arganda:2008jj,Guerrero:1997ku,Pich:2001pj,Miranda:2018cpf}
\begin{eqnarray}\label{eq.Fpitau}
F_\pi^\tau(s)=\Big(1-\frac{G_\rho F_\rho s}{F^2}\frac{1}{s_{\rho}}\Big)
\times \exp\left[\frac{-s}{96\pi^{2}F^{2}}\left({\rm Re}\left[A[m_{\pi},M_{\rho},s]+\frac{1}{2}A[m_{K},M_{\rho},s]\right]\right)\right]\,.
\end{eqnarray}
The function
\begin{eqnarray} \label{eq:func2}
A\left(m_{P},\mu,s\right) =  \ln\left(m_{P}^{2}/\mu^{2}\right)+\frac{8m_{P}^{2}}{s}-\frac{5}{3}+\sigma_{P}^{3}\ln\left(\frac{\sigma_{P}+1}{\sigma_{P}-1}\right) \, .
\end{eqnarray}
To incorporate isospin-breaking effects, one approach is to multiply
$|F_\pi^\tau(s)|^2$ by the factor $S_{EW}G_{EM}(s)$, where
$S_{EW}=1.0233$ corresponds to the short distance correction~\cite{Davier03}. Additionally, $G_{EM}(s)$ accounts for the long-distance radiative correction, as described in~\cite{Flores}. Specifically, in our fit of the $\tau\rightarrow \nu_{\tau}2\pi$ decay data we
perform the following substitution
\bea |F_\pi^\tau(s)|^2\Rightarrow
S_{EW}G_{EM}(s)|F_\pi^\tau(s)|^2. \eea

The pion form-factor in $e^+e^-$ annihilation reads
\begin{eqnarray}
F_\pi^{ee}(s)&=\Big(1-\frac{G_\rho F_\rho
s}{F^2}\frac{1}{s_{\rho}}-\frac{G_\rho F_\omega
s}{3F^2}\frac{1}{s_\omega}\Pi_{\rho\omega}\frac{1}{s_{\rho}}
-\frac{4\sqrt{2}aB_0F_\omega(m_u-m_d)s}{3F^2}\frac{1}{s_\omega}\Big) \nonumber\\ &
\times \exp\left[\frac{-s}{96\pi^{2}F^{2}}\left({\rm Re}\left[A[m_{\pi},M_{\rho},s]+\frac{1}{2}A[m_{K},M_{\rho},s]\right]\right)\right]\,.
\label{eqformfactoree}
\end{eqnarray}
As defined in the appendix~\ref{appendix-pipiloop}, the parameter $a$
is associated with the combined coupling constant of the direct $\omega\pi\pi$ interaction.
In the first bracket of Eq.~\eqref{eqformfactoree}, the second term corresponds to the contribution from the $\rho\pi\pi$ coupling, the third term represents the contribution of $\rho-\omega$ mixing, and the
fourth term corresponds to contribution from the direct isospin-breaking coupling of $\omega$ to the pion pair.

The leading order contribution of the $\pi\pi(\gamma)$ intermediate state to the anomalous magnetic moment of the muon is given by~\cite{Gourdin:1969dm}
\begin{equation}
a_{\mu}^{\pi\pi(\gamma),\text{LO} }=\left(\frac{\alpha m_{\mu}}{3\pi}\right)^{2}\int_{4 m_\pi^2}^{\infty}\mathrm{d}s\frac{\hat{K}(s)}{s^{2}}R_{\pi\pi(\gamma)}(s)\text{\ ,}\label{eq:amu}
\end{equation}
where
\begin{eqnarray} \label{eq:amusup}
  R_{\pi\pi(\gamma)}(s)=\frac{3s}{4\pi\alpha^{2}} \, \sigma^{(0)}\left(e^{+}e^{-}\rightarrow \pi\pi(\gamma)\right)\ ,
\end{eqnarray}
and the kernel function is defined as follows:
\begin{equation} \label{eq:amuker}
\hat{K}(s)=\frac{3s}{m_{\mu}^{2}}\left.\left[\frac{\left(1+x^{2}\right)(1+x)^{2}}{x^{2}}\left(\ln(1+x)-x+\frac{x^{2}}{2}\right)\right.\left.+\frac{x^{2}}{2}\left(2-x^{2}\right)+\frac{1+x}{1-x}x^{2}\ln x\right]\text{\ ,}\right.
\end{equation}
with
\begin{eqnarray} \label{eq:amusup2}
x=\frac{1-\beta_{\mu}(s)}{1+\beta_{\mu}(s)}, & \qquad & \qquad \beta_{\mu}(s)=\sqrt{1-\frac{4m_{\mu}^{2}}{s}}\ .
\end{eqnarray}
Note that in the formula for $a_{\mu}^{\pi\pi(\gamma),\text{LO} }$ in Eq.~\eqref{eq:amu}, the integration is performed from 4$m_\pi^2$
to $\infty$. In this work, we focus on the momentum dependence of $\rho-\omega$ mixing. Therefore, we only describe the pion vector form factor up to 900 MeV. To extend the study by considering higher energies, we must consider the effects of the excited resonances, such as $\rho^\prime(1450)$ and $\rho^{\prime\prime}(1700)$, etc. However, these are beyond the scope of this work. It is interesting to note that the $1/s^2$ enhancement factor in Eq.~\eqref{eq:amu} gives higher weight to the lowest
lying resonance $\rho(770)$ that couples strongly to $\pi^+\pi^-$.

The bare cross section including final-state radiation takes the following form~\cite{Gluza:2002ui,Czyz:2004rj,Bystritskiy:2005ib,Colangelo:2018mtw}
\begin{eqnarray}
\sigma^{(0)}(e^+e^-\to\gamma^*\to\pi^+\pi^-(\gamma)) = \Big[ 1 + \frac{\alpha}{\pi} \eta(s) \Big]  \frac{\pi \big|\alpha(s)\big|^2}{3s} \sigma_\pi^3(s) \big| F_\pi^{ee}(s) \big|^2 \frac{s+2m_e^2}{s \sigma_e(s)},
\end{eqnarray}
where
\begin{align}
\label{FSR}
	\begin{split}
			\eta(s) &= \frac{3(1+\sigma_\pi^2(s))}{2\sigma_\pi^2(s)} - 4 \log\sigma_\pi(s) + 6 \log \frac{1+\sigma_\pi(s)}{2} + \frac{1+\sigma_\pi^2(s)}{\sigma_\pi(s)} F(\sigma_\pi(s)) \\
				&\quad - \frac{(1-\sigma_\pi(s))\big(3+3\sigma_\pi(s)-7\sigma_\pi^2(s)+5\sigma_\pi^3(s)\big)}{4\sigma_\pi^3(s)} \log\frac{1+\sigma_\pi(s)}{1-\sigma_\pi(s)}, \\
			F(x) &= -4\dilog(x)+4\dilog(-x)+2\log x \log\frac{1+x}{1-x} + 3 \dilog\Big(\frac{1+x}{2} \Big) - 3\dilog\Big(\frac{1-x}{2} \Big) + \frac{\pi^2}{2}, \\
			\dilog(x) &= - \int_0^x dt \frac{\log(1-t)}{t}.
	\end{split}
\end{align}

The experimental data considered in this work are the pion form
factor $F_\pi^{ee}(s)$ of the $e^+e^-\rightarrow \pi^+\pi^-$
process measured by the OLYA~\cite{OLYACMD}, CMD~\cite{CMD2007}, BaBar~\cite{BaBar:2012bdw}, BESIII~\cite{BESIII}, KLOE~\cite{KLOE-2:2017fda}, CLEO~\cite{Xiao:2017dqv}, and SND~\cite{SND:2020nwa} Collaborations, the form
factor $F_\pi^\tau(s)$ of $\tau\rightarrow
\nu_{\tau}2\pi$ decay measured by the ALEPH~\cite{ALEPH} and CLEO~\cite{CLEO} Collaborations, and the decay width of $\omega \rightarrow
\pi^+\pi^-$~\cite{ParticleDataGroup:2020ssz}.
Note that in the experimentally published form factor data $F_\pi^{ee}(s)$, the vacuum polarization effects have been excluded through the subtraction of the hadronic running of $\alpha(s)$, and thus in our fitting of the form factor data $F_\pi^{ee}(s)$ the one-photon-reducible Fig.~\ref{FeynmanDiagram}(b) should not be considered.
Since we focus on the analysis of the $\rho-\omega$ mixing effect, we only take into account the form
factors $F_\pi^{ee}(s)$ and $F_\pi^{\tau}(s)$ data in the energy region of
600$\sim$900 MeV. Note that for the pion form
factor $F_\pi^{ee}(s)$, a tension between the two most precise measurements by BaBar and KLOE is observed in at and above the $\rho$ peak region, while the other measurements are consistent with theirs within the given uncertainty. To demonstrate the impact of the momentum dependence of $\rho-\omega$ mixing and to avoid the tension between the BaBar and KLOE data, we conduct four separate fits. Specifically, in Fits~Ia and Ib, we fit all data sets but BaBar with momentum-independent $\Pi_{\rho\omega}$ and momentum-dependent $\Pi_{\rho\omega}$, respectively.
In Fits~IIa and IIb, we fit all data sets but KLOE with momentum-independent $\Pi_{\rho\omega}$ and momentum-dependent $\Pi_{\rho\omega}$, respectively.
Fits Ia and IIa involve eight free parameters: $M_\rho$, $G_\rho$, $F_\rho$, $M_\omega$, $\Gamma_\omega$, $a$, the real and imaginary part of constant $\Pi_{\rho\omega}$. There are nine free parameters in Fits~Ib and~IIb: $M_\rho$, $G_\rho$, $F_\rho$, $M_\omega$, $\Gamma_\omega$, $a$, $X_W^r$, $X_Z^r$, and $X_R^r$.
As defined in the appendix~\ref{appendix-loop-contribution}, $X_W^r$, $X_Z^r$, and $X_R^r$ are the corresponding parameters for the counterterms.

\begin{figure*}[t]
\centering
\includegraphics[height=15cm,width=18cm]{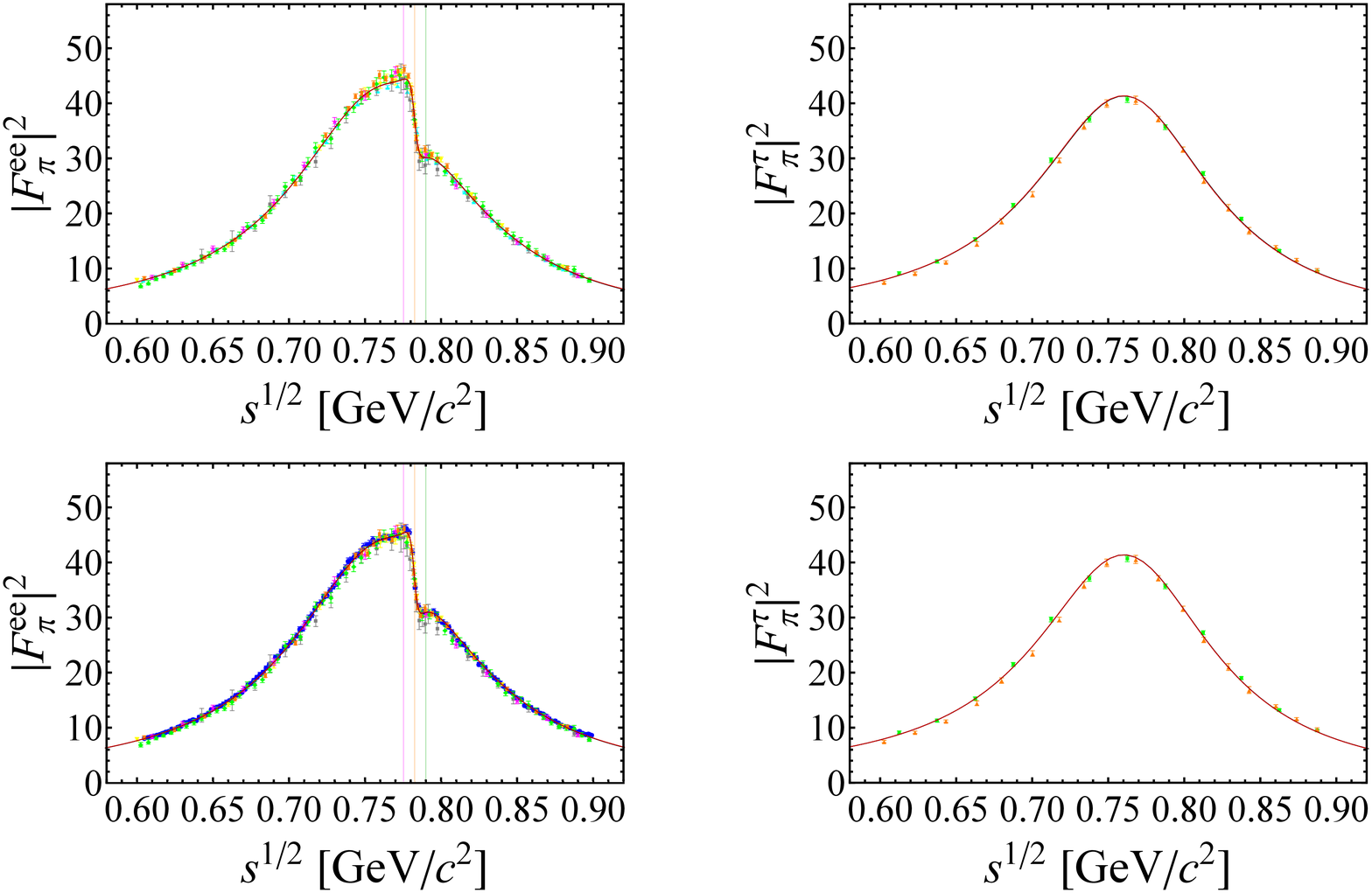}
\caption{  Fit results of the pion form factor in the $e^+e^-
\rightarrow \pi^+\pi^-$ process (left panel) and the $\tau\rightarrow
\nu_\tau 2\pi$ process (right panel) in the energy region of
600$\sim$900 MeV. The data of $e^+e^-$
annihilation are taken from the the OLYA (Gray)~\cite{OLYACMD}, CMD (Yellow)~\cite{CMD2007}, BaBar (Blue)~\cite{BaBar:2012bdw}, BESIII (Green)~\cite{BESIII}, KLOE (Cyan)~\cite{KLOE-2:2017fda}, CLEO (Magenta)~\cite{Xiao:2017dqv}, and SND (Orange)~\cite{SND:2020nwa} Collaborations. The $\tau$ decay data are taken from the ALEPH (Orange)~\cite{ALEPH} and
CLEO (Green)~\cite{CLEO} collaborations. Fits~Ia and Ib fit all data sets but BaBar (top), Fits~IIa and IIb fit all data sets but KLOE (bottom). Fits~Ia and IIa use momentum-independent $\Pi_{\rho\omega}$ and are shown as the red dashed lines. Fits~Ib and IIb use momentum-dependent $\Pi_{\rho\omega}$ and are shown as the black solid lines. The vertical lines lie at $M_\rho$, $M_\omega$, and $2M_\omega-M_\rho$ (from left to right).
} \label{fig2}
\end{figure*}

\begin{table}
\caption{\label{table1} The fit results of the parameters. Fits~Ia and Ib fit all data sets but BaBar, Fits~IIa and IIb fit all data sets but KLOE. Fits~Ia and IIa use momentum-independent $\Pi_{\rho\omega}$, while Fits~Ib and IIb use momentum-dependent $\Pi_{\rho\omega}$.}
\renewcommand{\arraystretch}{1.2}
\begin{center}
\begin{tabular}{l|cccc}
\toprule
         & Fit~Ia
         & Fit~Ib
         & Fit~IIa
         & Fit~IIb
  \\
\hline
$M_\rho$ [MeV]     &    $ 775.35 \pm 0.10 $ &   $ 775.68 \pm 0.12 $   &    $ 775.45\pm0.10 $&    $ 775.55\pm0.11 $     \\
$G_\rho$ [MeV]     &    $ 55.25 \pm 0.09 $ &   $ 55.74 \pm 0.08 $   &    $ 54.21\pm0.09 $&    $ 55.03\pm0.07 $     \\
$F_\rho$ [MeV]     &    $152.65\pm 0.29 $&     $151.40\pm 0.21 $    &    $ 155.65\pm0.23 $&    $ 153.38\pm0.31 $   \\
$M_\omega$ [MeV]     &    $ 782.59 \pm 0.13 $ &   $ 782.68 \pm 0.12 $   &    $ 782.39\pm0.11 $&    $ 782.45\pm0.11 $     \\
$\Gamma_\omega$ [MeV]     &    $ 8.97 \pm 0.27 $ &   $ 9.03 \pm 0.26 $   &    $ 8.04\pm0.16 $&    $ 8.16\pm0.17 $     \\
$a~[\text{GeV}^{-1}]$        &  $-0.0020\pm 0.0150$ &    $-0.0054\pm 0.0010$  &    $ -0.1066\pm0.0152 $&    $ -0.0067\pm0.0009 $   \\
$\text{Re}(\Pi_{\rho\omega}) [\text{MeV}^2]$    & $ -3372 \pm 112 $ &  -  &    $ -3799 \pm 85 $ &    -   \\
$\text{Im}(\Pi_{\rho\omega}) [\text{MeV}^2]$    & $296\pm 669 $ &  -  &   $ -4544 \pm 704 $ &    -  \\
$X_W^r~[\text{GeV}^{-6}]$    &  - &    $-0.141\pm 0.013$     &    -&    $ -0.177\pm0.008 $    \\
$X_Z^r~[\text{GeV}^{-4}]$    &  - &   $0.195\pm 0.016$   &    -&    $ 0.303\pm0.007 $  \\
$X_R^r~[\text{GeV}^{-2}]$    &  - &  $-0.081\pm 0.006$    &    -&    $ -0.133\pm0.003 $   \\
\hline
 ${\chi^2}/{\rm d.o.f.}$ &  $\frac{410.7}{(238-8)}=1.79$  &  $\frac{405.8}{(238-9)}=1.77$    &  $\frac{392.2}{(341-8)}=1.18$ &  $\frac{394.6}{(341-9)}=1.19$     \\
\hline
 $a_\mu^{\pi\pi}|_{[0.6,0.9] \text{GeV}} [\times 10^{10}]$ &  $367.72\pm 1.07$  &  $367.80\pm 2.92$    &  $375.41\pm 1.03$ &  $375.29\pm2.21$      \\
\botrule
\end{tabular}
\end{center}
\renewcommand{\arraystretch}{1.0}
\end{table}

In Fig.~\ref{fig2}, the
fitted results of the fits using momentum-independent $\Pi_{\rho\omega}$ (Fits~Ia and IIa) and momentum-dependent $\Pi_{\rho\omega}$ (Fits~Ib and IIb) are shown as red dotted lines and black solid lines, respectively.
The fitted parameters as well as the $\chi^2/\text{d.o.f.}$ are given in Table~\ref{table1}.
It is intriguing to compare the results obtained from fits utilising momentum-independent $\Pi_{\rho\omega}$ and momentum-dependent $\Pi_{\rho\omega}$ for the same datasets. Specifically, we compare Fit~Ia and Ib, Fit~IIa and IIb, respectively.
When examining the pion form factors $|F_\pi^{ee}(s)|^2$ and $|F_\pi^{\tau}(s)|^2$, we observe that the differences between the theoretical predictions of the fits using momentum-independent $\Pi_{\rho\omega}$ and the corresponding ones using momentum-dependent $\Pi_{\rho\omega}$ are tiny.
Furthermore, one notes that for the pion form factor $|F_\pi^{ee}(s)|^2$ in Fits~Ia and Ib, the theoretical predictions are much higher than the KLOE data at the $\rho$ peak, and these deviations contribute a lot to their value of $\chi^2$. Thus we conclude that both the momentum-independent $\Pi_{\rho\omega}$ and momentum-dependent $\Pi_{\rho\omega}$ can describe the
data well, and the discordances among different collaborations contribute significantly to $\chi^2$ values in the fits.

In the last line of Table~\ref{table1}, we give the results of $a_\mu^{\text{HVP,LO}}[\pi^+\pi^-]$, evaluated between 0.6 GeV and 0.9 GeV. The differences between the results using the momentum-independent $\Pi_{\rho\omega}$ and the results using the momentum-dependent $\Pi_{\rho\omega}$ for the same datasets, namely the differences between Fits~Ia and Ib, Fits~IIa and IIb, respectively, are negligible.

\begin{figure*}[t]
\centering
\includegraphics[width=\linewidth]{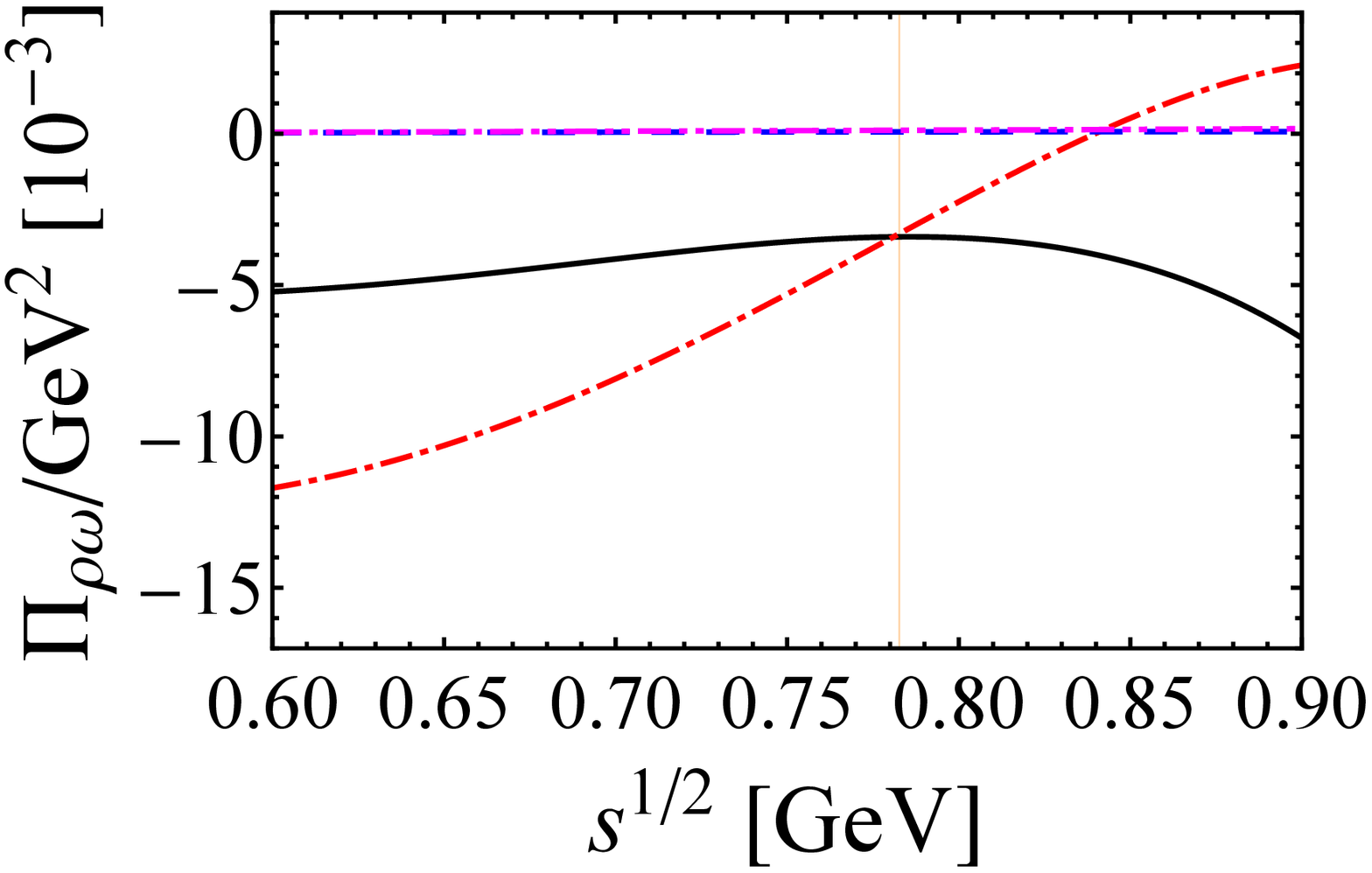}
\caption{ The momentum dependence of the mixing amplitudes
$\Pi_{\rho\omega}(s)$. The black solid and red dot-dashed
lines correspond to the real part of $\Pi_{\rho\omega}(s)$ in Fits~Ib and IIb, respectively.
The and blue dashed and magenta dash-dot-dotted lines correspond to the imaginary part of $\Pi_{\rho\omega}(s)$ in Fits~Ib and IIb, respectively.
The vertical line lies at $M_\omega$.} \label{fig_Pirhoomega}
\end{figure*}

In Fig.~\ref{fig_Pirhoomega}, we plot the real and imaginary parts of the
mixing amplitudes $\Pi_{\rho\omega}(s)$ in Fits~Ib and IIb. It is found that the real
part is dominant within the $\rho-\omega$ mixing region.
The real part in Fit IIb demonstrates a significant momentum dependence, whereas the real part in Fit Ib displays a smooth momentum dependence.
Additionally, it should be noted that the real parts of the two fits nearly reach the same point at $s^{1/2}=M_\omega$.
In comparison to the real part, the
imaginary part is rather small.
At $s=M_\omega^2$,
in Fit~Ib the mixing amplitude $\Pi_{\rho\omega}(M_\omega^2)=(-3405.0+62.1i)$ MeV$^2$, and in Fit~IIb $\Pi_{\rho\omega}(M_\omega^2)=(-3316.3+113.7i)$ MeV$^2$.
The smallness of the imaginary part is
consistent with the findings reported in Refs.~\cite{Renard,Connell98},
though therein the effect of direct $\omega_I \rightarrow
\pi^+\pi^-$ was not considered.
It is worth mentioning that larger imaginary part is obtained
in~\cite{Mitchell,MLYan00} by using global color model and a chiral
constituent quark model, respectively.
By utilising our fitted parameter results, we proceed to calculate the ratio of the two-pion couplings associated with the isospin-pure $\omega$ and $\rho$
\bea
G=\frac{g_{\omega_I\pi\pi}}{g_{\rho_I\pi\pi}}=\frac{4\sqrt{2}a B_0(m_u-m_d)}{G_\rho}.
\eea
The results are $G=(2.1\pm1.1)\times 10^{-3}$ in Fit~Ib, and $G=(2.6\pm1.2)\times 10^{-3}$ in Fit~IIb.
Note the value of $G$ is expected to be of order $\alpha=1/137$ in Ref.~\cite{Renard}.
The central values of our results of $G$ agree with the expectation in Ref.~\cite{Renard}, while they are lower than other two estimations, namely $G=(5.0\pm1.7_{\text{model}}\pm 1.0_{\text{data}})\times 10^{-2}$ in~\cite{Wolfe:2010gf}, and $G=(3.47\pm0.64)\times 10^{-2}$ in~\cite{Benayoun:2012etq}.
As shown in Table~\ref{table1}, the differences of ${\chi^2}/{\rm d.o.f.}$ between the momentum-dependent fits and the momentum-independent fits for the same data sets are tiny.
The $\chi^2$ of Fit IIa is slightly lower than the $\chi^2$ of Fit IIb, though Fit IIa contains one less fitting parameter than Fit IIb. One observes that the magnitude of the imaginary part of $\Pi_{\rho\omega}$ in Fit IIa is significantly greater than those in other three fits.
In our framework, the imaginary part of $\Pi_{\rho\omega}$ arises from the $\pi^0\gamma$ and $\pi\pi$
real intermediate states.
By considering the decay widths of $\omega \rightarrow \pi^0\gamma$ and $\rho \rightarrow \pi^0\gamma$, the imaginary part of $\Pi_{\rho\omega}$ contributed from the $\pi^0\gamma$ intermediate state can be estimated to be approximately $-$150 MeV$^2$~\cite{Renard,Colangelo:2022prz}.
If one use the estimated ratio of the two-pion couplings of the isospin-pure $\omega$ and $\rho$: $G \sim \alpha=1/137$~\cite{Renard}, one can obtain that the $\pi\pi$ intermediate state contribution to the imaginary part of $\Pi_{\rho\omega}$ to be in the order of several hundred MeV$^2$. In our momentum-dependent scheme, the imaginary part of $\Pi_{\rho\omega}$ due to the $\pi^0\gamma$ and $\pi\pi$ intermediate states are explicitly computed, and both the numerical results of Im$\Pi_{\rho\omega}$ in Fits Ib and IIb are of the order of one hundred MeV$^2$ as expected. While in the momentum-independent Fits Ia and IIa, the imaginary part of $\Pi_{\rho\omega}$ is a free fitting parameter. As can be seen in Table~\ref{table1}, the fitted results of Im$\Pi_{\rho\omega}$ and the parameter ``$a$'' in Fit IIa are unreasonably large.
The fitted results of Im$\Pi_{\rho\omega}$ and the parameter ``$a$'' in Fit Ia exhibit large error bars. Consequently, we conclude that though both the momentum-independent and momentum-dependent $\rho-\omega$ mixing schemes can describe the $e^+e^- \rightarrow \pi^+\pi^-$ data well, the momentum-dependent $\rho-\omega$ mixing scheme is more self-consistent, especially given the reasonable imaginary part of $\Pi_{\rho\omega}$ extracted.

We would like to point out that the direct $\omega_I\rightarrow \pi^+\pi^-$ coupling is generally an unknown quantity, and it impacts $F_\pi^{ee}(s)$ in two ways, both through the third term in the first bracket of Eq.~\eqref{eqformfactoree} as it appearing as real intermediate state in the contributions to $\Pi_{\rho\omega}$ and through the fourth term in that bracket.
Conventionally, the direct $\omega_I\rightarrow \pi^+\pi^-$ is assumed to cancel out in $e^+e^- \rightarrow
\pi^+\pi^-$ due to the fact that $\omega$ and $\rho$ are quasidegenerate and that the 2$\pi$ channel dominates the $\rho$ decay~\cite{Renard}. While theoretical models that do not neglect the direct $\omega_I\rightarrow \pi^+\pi^-$ coupling may be more comprehensive, especially given the availability of high-precision data nowadays.
Note that in Refs.~\cite{Colangelo:2018mtw,Colangelo:2022prz} the pion form factor has been studied in a model-independent way using dispersion theory. The $\rho-\omega$ mixing is subsumed in one parameter $\epsilon_\omega$, which should contain a small imaginary part originating from the radiative intermediate states (with an estimated phase of approximately 4 degrees). Note since the direct $\omega_I\rightarrow \pi^+\pi^-$ coupling is not considered in Refs.~\cite{Colangelo:2018mtw,Colangelo:2022prz}, the $\epsilon_\omega$ term there is actually a combination of the $\rho-\omega$ mixing and direct $\omega_I\rightarrow \pi^+\pi^-$. Therefore, it cannot be directly compared to the $\Pi_{\rho\omega}$ discussed in this context. (At $s=M_\omega^2$, our $\Pi_{\rho\omega}(M_\omega^2)$ in Fits~Ib and IIb contains negative phase.) Nevertheless, the ratio between the on-$\omega$-mass-shell $\gamma^\ast \rightarrow \omega \rightarrow \pi\pi$ transition amplitude and $\gamma^\ast \rightarrow \rho \rightarrow \pi\pi$ transition amplitude (without $\pi\pi$ final state interaction) should be model independent. With $s=M_\omega^2$, the ratio between the second term and the first term in Eq.~(2.5) of~\cite{Colangelo:2022prz} yields $R_{\omega\rho}=\text{Amplitude}({\gamma^\ast \rightarrow \omega \rightarrow \pi\pi})/\text{Amplitude}({\gamma^\ast \rightarrow \rho \rightarrow \pi\pi})=(0.178 \pm 0.003)\times e^{i(4.66\pm 1.13)^\circ}$, using Re$\epsilon_\omega=(1.97\pm0.03)\times 10^{-3}$ and $\delta_\epsilon=(4.5\pm 1.2)^\circ$ obtained therein. One observes that the difference between the phase of $R_{\rho\omega}(M_\omega^2)$ and $\delta_\epsilon$ is tiny.
In the present work, the ratio between the sum of the third term and fourth terms and the sum of
the first and second terms in the first bracket of Eq.~\eqref{eqformfactoree} predicts $R_{\omega\rho}=(0.155 \pm 0.002)\times e^{i(5.80\pm 1.71)^\circ}$ and $R_{\omega\rho}=(0.150 \pm 0.002)\times e^{i(3.67\pm 1.71)^\circ}$ in Fit Ib and IIb, respectively. One can see that our results of $R_{\omega\rho}$ roughly agree with that in~\cite{Colangelo:2022prz}.

Using the central values of the
fitted parameters of our best fit (Fit~IIb) in Table~\ref{table1}, we calculate the decay width of
$\omega \rightarrow \pi^+\pi^-$
\begin{eqnarray}   \label{eq.Omegapipi}
\Gamma_{\omega \rightarrow \pi^+\pi^-}&=&\frac{1}{192\pi
F^4}(M_\omega^2-4m_\pi^2)^{\frac{3}{2}}  \times \Big(\exp\left[\frac{-M_\omega^2}{96\pi^{2}F^{2}}\left({\rm Re}\left[A[m_{\pi},M_{\rho},M_\omega^2]+\frac{1}{2}A[m_{K},M_{\rho},M_\omega^2]\right]\right)\right]\Big)^2 \nonumber\\
&& \times \Big|
8\sqrt{2}B_0(m_u-m_d)a+\frac{2G_\rho
\Pi_{\rho\omega}(M_\omega^2)}{M_\omega^2-M_\rho^2-i(M_\omega
\Gamma_\omega-M_\rho \Gamma_\rho)}\Big|^2\nonumber\\
 &=&0.013\mid
(0.29)+(-0.22+3.35i)\mid^2 \,.
\end{eqnarray}
From Eq.~\eqref{eq.Omegapipi}, we can find that the first term due
to the direct $\omega_I \rightarrow \pi^+\pi^-$ is smaller than the
second term due to the $\rho-\omega$ mixing by one order of magnitude. Within 1$\sigma$ uncertainties,
our theoretical value of the branching fraction is
$\mathscr{B}(\omega \rightarrow \pi^+\pi^-)=(1.48\pm 0.10)\times
10^{-2}$, which is consistent with the values provided in PDG~\cite{ParticleDataGroup:2020ssz} and by the recent dispersive
analysis~\cite{Kubis2017}.

Regarding the mass of $\omega$ meson, previous studies~\cite{BaBar:2012bdw,Colangelo:2018mtw,Colangelo:2022prz} have pointed out that the result extracted from $e^+e^- \rightarrow
\pi^+\pi^-$ is substantially lower than the current PDG average~\cite{ParticleDataGroup:2020ssz}, which primarily relies on $e^+e^- \rightarrow 3\pi$ and $e^+e^- \rightarrow \pi^0\gamma$ experiments.
The discrepancy amounts to approximately 1 MeV, corresponding to around 5 $\sigma$ considering the current precision.
It has been observed that the fitted value for $M_\omega$ and the phase of $\epsilon_\omega$ are strongly correlated~\cite{BaBar:2012bdw,Colangelo:2018mtw,Colangelo:2022prz}.
Note that the direct $\omega_I\rightarrow \pi^+\pi^-$ coupling has not been considered in~\cite{BaBar:2012bdw,Colangelo:2018mtw,Colangelo:2022prz}.
As indicated in Table~\ref{table1} above, our fitted results for the mass of $\omega$ agree well with the value in PDG: $M_\omega= 782.66 \pm 0.13 $ MeV, and this agreement remains unaffected by the inclusion or exclusion of the momentum dependence of $\Pi_{\rho\omega}$.
We also find that strong correlation (80\%) exists between the parameter ``$a$'', which quantify the direct $\omega_I\rightarrow \pi^+\pi^-$ coupling, and the mass of $\omega$. As mentioned earlier, the direct $\omega_I\rightarrow \pi^+\pi^-$ coupling influences both the imaginary part and real part of the amplitude and thus affects the phase of $R_{\rho\omega}(M_\omega^2)$.
Note that the phase of $R_{\rho\omega}(M_\omega^2)$ roughly agrees with the phase of $\epsilon_\omega$.
Thus our observations align with with those in Refs.~\cite{BaBar:2012bdw,Colangelo:2018mtw,Colangelo:2022prz}, namely strong correlation exists
between the mass of the omega meson and the phase of $\epsilon_\omega$.
Our findings suggest that the inclusion of direct $\omega_I\rightarrow \pi^+\pi^-$ coupling is likely crucial in the analysis aimed at extracting the $\omega$ mass from the $e^+e^- \rightarrow
\pi^+\pi^-$ process.

\section{Conclusions}  \label{section.Conclusions}

We have used the resonance chiral theory to study the $\rho-\omega$ mixing. In particular, we have analyzed the impact of the
momentum dependence of the $\rho-\omega$ mixing on the describing the pion vector form factor in the
$e^+e^-\rightarrow \pi^+\pi^-$ process and its contribution to the anomalous magnetic moment of the muon $a_\mu$.
The incorporation of momentum dependence arises from the calculation of loop contributions, which corresponds to the next-to-leading orders in $1/N_C$ expansion.
Through fitting to the data of $e^+e^-\rightarrow \pi^+\pi^-$ and and $\tau\rightarrow
\nu_{\tau}2\pi$ processes
within the energy range of 600$\sim$900 MeV and the decay width of $\omega \rightarrow \pi^+\pi^-$,
we find that the $\rho-\omega$ mixing amplitude is dominated by its real part, and its imaginary part is relatively small.
Although both momentum-independent and momentum-dependent $\rho-\omega$ mixing schemes yield satisfactory data descriptions, the latter proves to be more self-consistent due to the reasonable imaginary part of the mixing matrix element $\Pi_{\rho\omega}$.
Regarding the contribution to anomalous magnetic moment of the muon
$a_\mu^{\pi\pi}|_{[0.6,0.9]\text{GeV}}$, the results obtained from fits considering the momentum-dependent $\rho-\omega$ mixing amplitude align well with those from corresponding fits that exclude the momentum dependence of $\rho-\omega$ mixing, within the margin of error.
Additionally, we provide the ratio of the isospin-pure $\omega$ and $\rho$ two-pion couplings, denoted as $G=g_{\omega_I\pi\pi}/g_{\rho_I\pi\pi}$, and observe that $\rho-\omega$ mixing plays a crucial role in the decay width of $\omega \rightarrow \pi^+\pi^-$. Furthermore, we ascertain that including the direct $\omega_I\rightarrow \pi^+\pi^-$ coupling is essential in analysing the extraction of the mass of the $\omega$ meson from the $e^+e^- \rightarrow \pi^+\pi^-$ process.

\section*{Acknowledgments}

We are grateful to Pablo Roig for helpful discussions and valuable suggestions.
This work is supported in part by the Fundamental Research Funds
for the Central Universities under Grants No.~FRF-BR-19-001A, by the National Natural Science Foundation of China (NSFC) under Grants No.~11975028, No.~11974043.

\appendix

\section{ Loop contributions}\label{appendix-loop-contribution}

\subsection{Diagram (c): kaon loops}\label{appendix-kaonloop}

Using the $\rho K\bar{K}$ and $\omega K\bar{K}$ vertexes given by the Lagrangian in Eq.~\eqref{eq.interactionRchiT}:
$ iG_V/\sqrt{2}\langle V_{\mu\nu}u^\mu u^\nu\rangle= iG_V/F^2 \rho^0_{\mu\nu}(\partial^\mu K^+ \partial^\nu K^- - \partial^\mu K^0 \partial^\nu \bar{K}^0)
+iG_V/F^2 \omega_{\mu\nu}(\partial^\mu K^+ \partial^\nu K^- + \partial^\mu K^0 \partial^\nu \bar{K}^0)+...$, we can calculate the charged and the neutral kaon loops contribution to the amplitude
 \begin{eqnarray}
\Pi_{\rho\omega}^{\text{kaon,charged}}
&=&
\frac{G_V^2 p^4}{192F^4\pi^2
}
\bigg\{(1-\frac{6m_{K^+}^2}{p^2})(\lambda_{\infty}-\ln\frac{m_{K^+}^2}{\mu^2})\nonumber\\
&&+\frac{5}{3}-\frac{8m_{K^+}^2}{p^2}
-\sigma_{K^+}^3\ln(\frac{\sigma_{K^+}+1}{\sigma_{K^+}-1})\bigg\}\,,
\end{eqnarray}
and
 \begin{eqnarray}
\Pi_{\rho\omega}^{\text{kaon,neutral}}
&=&
-\frac{G_V^2 p^4}{192F^4\pi^2
}
\bigg\{(1-\frac{6m_{K^0}^2}{p^2})(\lambda_{\infty}-\ln\frac{m_{K^0}^2}{\mu^2})\nonumber\\
&&+\frac{5}{3}-\frac{8m_{K^0}^2}{p^2}
-\sigma_{K^0}^3\ln(\frac{\sigma_{K^0}+1}{\sigma_{K^0}-1})\bigg\}\,,
\end{eqnarray}
where $\sigma_P\equiv\sqrt{1-4m_P^2/p^2}$
and $\lambda_{\infty}\equiv \frac{1}{\epsilon}-\gamma_E+1+\ln4\pi$ with
$\epsilon=2-\frac{d}{2}$ and $\gamma_E$ being the Euler constant.

The non-vanishing of the structure function proceeds from the mass difference between the charged and neutral kaons as follows
 \begin{eqnarray}
S_{\rho\omega}^{(c)}=\frac{1}{m_u-m_d}(\Pi_{\rho\omega}^{\text{kaon,charged}}+\Pi_{\rho\omega}^{\text{kaon,neutral}})\,.
\end{eqnarray}

\subsection{Diagram (d): $\pi\pi$ loop}\label{appendix-pipiloop}

For the isospin-violating vertex of $\omega_I\rightarrow
\pi^+\pi^-$, we construct the Lagrangian
\begin{eqnarray}
\mathcal{L}_{\omega_I\rightarrow
\pi^+\pi^-}&=&a_1i\langle V_{\mu\nu}\{\chi_+, u^\mu
u^\nu\}\rangle +a_2i\langle V_{\mu\nu}u^\mu \chi_+
u^\nu\rangle \nonumber\\
&=&(a_1-\frac{1}{2}a_2)\frac{8\sqrt{2}B_0i}{F^2}\Delta_{ud}\,\omega_{\alpha\beta}\pi^{+\alpha}\pi^{-\beta}.
\end{eqnarray}
For convenience, we define the combination $a\equiv a_1-\frac{1}{2}a_2.$
The $\pi\pi$-loop contribution to the structure function can be calculated, which reads
 \begin{eqnarray}
S_{\rho\omega}^{(d)}
&=&
\frac{\sqrt{2}G_VB_0a}{12F^4\pi^2
}
p^4\bigg\{(1-\frac{6m_\pi^2}{p^2})(\lambda_{\infty}-\ln\frac{m_\pi^2}{\mu^2})\nonumber\\
&&\hspace{1.5cm}+\frac{5}{3}-\frac{8m_\pi^2}{p^2}
-\sigma_\pi^3\ln(\frac{\sigma_\pi+1}{\sigma_\pi-1})\bigg\}\,.
\end{eqnarray}

\subsection{Diagram (e): $\pi$-tadpole loop}\label{appendix-pitadpole}

According to the Lorentz, $P$ and $C$ invariances, the Lagrangian corresponding to the interaction of $\omega_I\rho_I \pi\pi$ can be constructed as follows:
\begin{eqnarray}\label{eq.OmegaRhoPP}
\mathcal{L}_{\omega_I\rho_I PP}&=& b_1\langle V_{\mu\nu}V^{\mu\nu}(u^\alpha
u_\alpha \chi_+ +\chi_+u^\alpha u_\alpha)\rangle \nonumber\\
&+&b_2\langle V_{\mu\nu}V^{\mu\nu}u^\alpha \chi_+u_\alpha \rangle +b_3\langle V_{\mu\nu}\chi_+V^{\mu\nu}u^\alpha u_\alpha \rangle \nonumber \\
&+&
b_4\langle V_{\mu\nu}u^\alpha V^{\mu\nu} (\chi_+u_\alpha+u_\alpha\chi_+) \rangle \nonumber \\
&+&b_5\langle V_{\mu\alpha}V^{\nu\alpha}u^\mu u_\nu\chi_+ +V^{\nu\alpha}V_{\mu\alpha}\chi_+ u_\nu u^\mu\rangle \nonumber\\
&+&b_6\langle V_{\mu\alpha}V^{\nu\alpha}u^\mu\chi_+ u_\nu\rangle +b_7\langle V_{\mu\alpha}\chi_+V^{\nu\alpha}u^\mu u_\nu\rangle\nonumber \\
&+& b_8\langle V_{\mu\alpha}V^{\nu\alpha}u_\nu u^\mu \chi_+
+V^{\nu\alpha}V_{\mu\alpha}\chi_+  u^\mu u_\nu\rangle \nonumber \\
&+&b_9\langle V_{\mu\alpha}V^{\nu\alpha} u_\nu\chi_+u^\mu\rangle
+b_{10}\langle V_{\mu\alpha}\chi_+V^{\nu\alpha}u_\nu u^\mu \rangle \nonumber \\
&+&b_{11}\langle V_{\mu\alpha}u^\alpha V^{\mu\beta}u_\beta  \chi_+
+V^{\mu\beta}u^\alpha V_{\mu\alpha} \chi_+u_\beta \rangle \nonumber\\
&+&b_{12}\langle V_{\mu\alpha}u^\alpha V^{\mu\beta}\chi_+u_\beta
+V^{\mu\beta}u^\alpha V_{\mu\alpha} u_\beta\chi_+ \rangle \nonumber \\
&+&b_{13}\langle V_{\mu\alpha}u_\beta V^{\mu\beta}u^\alpha  \chi_+
+V^{\mu\beta}u_\beta V_{\mu\alpha} \chi_+u^\alpha \rangle\nonumber\\
&+&b_{14}\langle V_{\mu\alpha}u_\beta V^{\mu\beta}\chi_+u^\alpha
+V^{\mu\beta}u_\beta V_{\mu\alpha}u^\alpha \chi_+ \rangle \nonumber \\
&+&g_1i\langle V_{\mu\nu}V^{\mu\nu}(u^\alpha \nabla_\alpha\chi_-+
\nabla_\alpha\chi_-u^\alpha)\rangle \nonumber\\
&+&g_2i\langle V_{\mu\nu}u^\alpha V^{\mu\nu} \nabla_\alpha\chi_-\rangle \nonumber \\
&+&g_3i\langle V_{\mu\beta}V^{\mu\alpha}u^\beta \nabla_\alpha\chi_-+
V^{\mu\alpha}V_{\mu\beta}\nabla_\alpha\chi_-u^\beta \rangle \nonumber \\
&+&g_4i\langle V_{\mu\beta}V^{\mu\alpha} \nabla_\alpha\chi_-u^\beta+
V^{\mu\alpha}V_{\mu\beta}u^\beta\nabla_\alpha\chi_- \rangle \nonumber \\
&+&g_5i\langle V_{\mu\beta}u^\beta V^{\mu\alpha} \nabla_\alpha\chi_-+
V^{\mu\alpha}u^\beta V_{\mu\beta}\nabla_\alpha\chi_- \rangle \nonumber\\
&+&\lambda_6^{VV}\langle V_{\mu\nu}V^{\mu\nu}\chi_+\rangle \,.
\end{eqnarray}
For simplicity, we define the combinations,
\begin{eqnarray}
h_1&\equiv&6b_1-b_2+3b_3+b_4-2g_1-g_2, \nonumber\\
h_2&\equiv&4b_5-b_6+3b_7+4b_8-b_9+3b_{10}+2b_{11}+2b_{12}\nonumber\\
&&+2b_{13}+2b_{14}-2g_3-2g_4-2g_5\,.
\end{eqnarray}

The mass difference between the charged and neutral pions in the internal lines of loops can be disregarded due to its higher-order magnitude beyond our scope of consideration. Consequently, the expanded expression of Lagrangian~\eqref{eq.OmegaRhoPP} can be simplified as follows:
\begin{eqnarray}
\mathcal{L}_{\omega_I\rho_I
\pi\pi}&=&\frac{4B_0}{F^2}h_1(m_u-m_d)\rho_{I\mu\nu}\omega^{\mu\nu}\pi_{\alpha}{\pi}^{\alpha}\nonumber\\
&-&\frac{2B_0}{F^2}\lambda_6^{VV}(m_u-m_d)\rho_{I\mu\nu}\omega^{\mu\nu}{\pi}^2\nonumber\\
&+&\frac{4B_0}{F^2}h_2(m_u-m_d)\rho_{I\mu\alpha}\omega^{\nu\alpha}\pi_{\mu}{\pi}^{\nu}\,.
\end{eqnarray}
With the above Lagrangian, the $\pi$-tadpole contribution to the $\rho-\omega$ mixing can be derived:
\begin{eqnarray}\label{eq:tadpoleloop}
S^{(e)}_{\rho\omega}&
=&-\frac{m_\pi^2B_0}{8\pi^2F^2}\bigg\{(-16\lambda_6^{VV}+4h_1 m_\pi^2+h_2 m_\pi^2)\nonumber\\
&&\hspace{2cm}\times(\lambda_\infty-\ln\frac{m_\pi^2}{\mu^2})+\frac{h_2}{2} m_\pi^2\bigg\}\,.
\end{eqnarray}

\subsection{Diagrams (f)-(i): $\pi^0\gamma$ loops}\label{appendix-pigammaloop}

In the loop diagrams (f)-(i), the resonance chiral effective Lagrangian describing
vector-photon-pseudoscalar (VJP) and vector-vector-pseudoscalar (VVP)
vertices has been given in
Ref.~\cite{Femenia}:
\begin{eqnarray} \label{eq.LagrangianVJP}
\mathcal{L}_{VJP}&=&\frac{c_1}{M_V}\epsilon_{\mu\nu\rho\sigma}\langle \{V^{\mu\nu},f_+^{\rho\alpha}\}\nabla_\alpha
u^\sigma\rangle \nonumber\\
&+&\frac{c_2}{M_V}\epsilon_{\mu\nu\rho\sigma}\langle \{V^{\mu\alpha},f_+^{\rho\sigma}\}\nabla_\alpha
u^\nu\rangle \nonumber\\
&+&\frac{ic_3}{M_V}\epsilon_{\mu\nu\rho\sigma}\langle \{V^{\mu\nu},f_+^{\rho\sigma}\}\chi_-\rangle \nonumber\\
&+&\frac{ic_4}{M_V}\epsilon_{\mu\nu\rho\sigma}\langle V^{\mu\nu}[f_-^{\rho\sigma},\chi_+]\rangle \nonumber\\
&+&\frac{c_5}{M_V}\epsilon_{\mu\nu\rho\sigma}\langle \{\nabla_\alpha
V^{\mu\nu},f_+^{\rho\alpha}\}u^\sigma\rangle \nonumber\\
&+&\frac{c_6}{M_V}\epsilon_{\mu\nu\rho\sigma}\langle \{\nabla_\alpha
V^{\mu\alpha},f_+^{\rho\sigma}\}u^\nu\rangle \nonumber\\
&+&\frac{c_7}{M_V}\epsilon_{\mu\nu\rho\sigma}\langle \{\nabla^\sigma
V^{\mu\nu},f_+^{\rho\alpha}\}u_\alpha\rangle \,,
\end{eqnarray}
and
\begin{eqnarray}  \label{eq.LagrangianVVP}
\mathcal{L}_{VVP}&=&d_{1}\epsilon_{\mu\nu\rho\sigma}\langle \{V^{\mu\nu},V^{\rho\alpha}\}\nabla_\alpha
u^\sigma\rangle \nonumber\\
&+&id_{2}\epsilon_{\mu\nu\rho\sigma}\langle \{V^{\mu\nu},V^{\rho\sigma}\}\chi_-\rangle \nonumber\\
&+&d_{3}\epsilon_{\mu\nu\rho\sigma}\langle \{\nabla_\alpha
V^{\mu\nu},V^{\rho\alpha}\}u^\sigma\rangle \nonumber\\
&+&d_{4}\epsilon_{\mu\nu\rho\sigma}\langle \{\nabla^\sigma
V^{\mu\nu},V^{\rho\alpha}\}u_\alpha\rangle \,.
\end{eqnarray}
The couplings involved, or their combinations, can be estimated by matching the leading operator product expansion of the $\langle
VVP\rangle $ Green function to the same quantity evaluated within R$\chi$T.
This procedure leads to high energy constraints on the resonance
couplings~\cite{Femenia}:
\begin{eqnarray} \label{eq.HighEnergyConstraints}
4c_3+c_1&=&0,\nonumber\\ c_1-c_2+c_5&=&0,\nonumber\\
c_5-c_6&=&\frac{N_c}{64\pi^2}\frac{M_V}{\sqrt{2}F_V},\nonumber\\
d_1+8d_2&=&-\frac{N_c}{64\pi^2}\frac{M_V^2}{F_V^2}+\frac{F^2}{4F_V^2},\nonumber\\
d_3&=&-\frac{N_c}{64\pi^2}\frac{M_V^2}{F_V^2}+\frac{F^2}{8F_V^2}.
\end{eqnarray}

Using the the effective vertices stated in Eqs.~\eqref{eq.LagrangianVJP} and ~\eqref{eq.LagrangianVVP}, the $\pi\gamma$ loop contribution, i.e., the summation of the loops diagrams (f)-(i), can be expressed as:
\begin{eqnarray}
i\Pi_{\rho\omega}\epsilon_{\rho\mu}\epsilon_\omega^\mu&=&
\frac{1}{p^2}\int
\frac{d^nk}{(2\pi)^n}\frac{-i}{k^2}\frac{i}{(p-k)^2-m_\pi^2}\nonumber\\
&\times&[(k\cdot
p)^2\epsilon_{\rho}^\mu\epsilon_{\omega\mu}-k^2p^2\epsilon_{\rho}^\mu\epsilon_{\omega\mu}
+p^2k\cdot\epsilon_{\rho}k\cdot\epsilon_{\omega}]\nonumber\\
&\times&\bigg\{\frac{-32e^2}{3 M_V^2F^2}\big[c_1(p-k)\cdot
k-c_2(p-k)\cdot p\nonumber\\
&&\hspace{0.2cm}-4c_3m_\pi^2-c_5p\cdot k+c_6p^2\big]^2\nonumber\\
&-&\frac{16\sqrt{2}F_Ve^2}{3
M_VF^2}\bigg[\frac{1}{M_\omega^2-k^2}+\frac{1}{M_\rho^2-k^2}\bigg]\nonumber\\
&\times&\big[d_1(p-k)^2+8d_2m_\pi^2+2d_3p\cdot k\big]
\nonumber\\
&\times&
\big[c_1(p-k)\cdot k-c_2(p-k)\cdot p-4c_3m_\pi^2\nonumber\\
&-&c_5p\cdot k+c_6p^2\big]-\frac{16F_V^2e^2}{3
F^2(M_\rho^2-k^2)(M_\omega^2-k^2)}\nonumber\\
&\times&\big[d_1(p-k)^2+8d_2m_\pi^2+2d_3p\cdot
k\big]^2 \bigg\}\,.\hspace{10.5cm}
\end{eqnarray}
The subsequent calculation is straightforward, while the result of the extracted electromagnetic structure function $E^{\pi\gamma}_{\rho\omega}\equiv E_{\rho\omega}^{(f)}+E_{\rho\omega}^{(g)}+E_{\rho\omega}^{(h)}+E_{\rho\omega}^{(i)}$ is too lengthy to be given here. Notes that in our numerical computation we employ the
high energy constraints in Eq.~\eqref{eq.HighEnergyConstraints} together with the fitted parameters given in
Ref.~\cite{chen2012}, and therefore all the parameters involved in $E^{\pi\gamma}_{\rho\omega}$ are known.

\subsection{ Counterterms and renormalized amplitude }\label{appendix-counterterms-renormalized-amp}

\begin{figure*}[ht]
\centering
\includegraphics[scale=0.6]{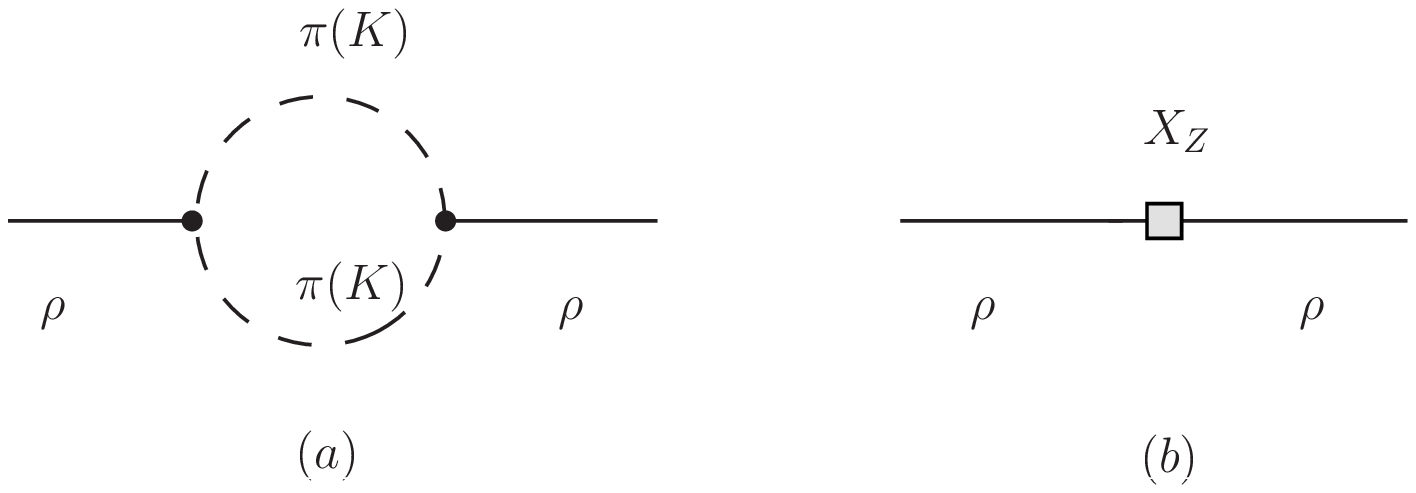}
\caption{ Feynman diagrams contributing to $\rho$ self-energy.
} \label{FeynmanDiagram_rho_selfenergy}
\end{figure*}

Since the $\omega$ meson predominantly decays into the three-pion state, its two-loop self energy diagram contributes beyond the NLO in $1/N_C$ and is not relevant for our current consideration.
The self-energy diagrams for the $\rho$ meson are depicted in Fig.~\ref{FeynmanDiagram_rho_selfenergy}. The Lagrangian needed to renormalize the $\rho$ meson one-loop self-energy has been given in Ref.~\cite{Rosell04},
\begin{eqnarray}\label{eq.XZ}
{\cal L}_{4Y}  &=&
\frac{X_{Y_1}}{2} \,\bra \nabla^2 V^{\mu\nu}
\left\{\nabla_\nu,\nabla^\sigma\right\} V_{\mu\sigma} \ket
 +
\frac{X_{Y_2}}{4} \,\bra \left\{\nabla_\nu,\nabla_\alpha\right\} V^{\mu\nu}
\left\{\nabla^\sigma,\nabla^\alpha\right\} V_{\mu\sigma} \ket \nonumber\\
&  + &
\frac{X_{Y_3}}{4} \,\bra \left\{\nabla^\sigma,\nabla^\alpha\right\} V^{\mu\nu}
\left\{\nabla_\nu,\nabla_\alpha\right\} V_{\mu\sigma} \ket\, .
\end{eqnarray}
Actually, only the combination of couplings $X_Y\equiv X_{Y_1}+X_{Y_2}+X_{Y_3}\equiv X_Y^r+\delta X_Y$
is relevant for this purpose. Using the Lagrangians in Eqs.~\eqref{eq.interactionRchiT} and~\eqref{eq.XZ}, the $\rho$ self-energy takes the form
 \begin{eqnarray}
\Sigma_{\rho}(p^2)
&=&
-\frac{G_V^2}{48F^4\pi^2
}
p^4\bigg\{(1-\frac{6m_\pi^2}{p^2})(\lambda_{\infty}-\ln\frac{m_\pi^2}{\mu^2})-\frac{8m_\pi^2}{p^2}
-\sigma_\pi^3\ln(\frac{\sigma_\pi+1}{\sigma_\pi-1})\nonumber\\
&&+(1-\frac{6m_K^2}{p^2})(\lambda_{\infty}-\ln\frac{m_K^2}{\mu^2})-\frac{8m_K^2}{p^2}
-\sigma_K^3\ln(\frac{\sigma_K+1}{\sigma_K-1})+\frac{10}{3}\bigg\}-X_Y p^4\,.
\end{eqnarray}
The renormalized $\rho$ mass fulfills
 \begin{eqnarray}
M_\rho^2=M_V^2+\Sigma_{\rho}(M_\rho^2)\,.
\end{eqnarray}
Since the physical $M_\rho$ is finite, one gets
 \begin{eqnarray}
\delta X_Y=-\frac{G_V^2}{48F^4 M_\rho^2\pi^2
}(1-6\frac{m_\pi^2}{M_\rho^2}-6\frac{m_K^2}{M_\rho^2})\lambda_{\infty}\,.
\end{eqnarray}
The wave-function renormalization constant of the $\rho$ meson is obtained from
 \begin{eqnarray}
Z_\rho=1+\frac{\partial \Sigma_\rho(p^2)}{\partial p^2}|_{p^2=M_\rho^2}\,.
\end{eqnarray}

In our calculation of the $\rho-\omega$ mixing, the tree amplitudes can only absorb the ultraviolet divergence that is proportional to $p^0$. To cancel the $O(p^2)$, $O(p^4)$ and $O(p^6)$ ultraviolet divergence originating from the loop contribution $S_{\rho\omega}^{(c)}$, $S_{\rho\omega}^{(d)}$, $S_{\rho\omega}^{(e)}$, and $E_{\rho\omega}^{\pi\gamma}$, we construct the counterterms as follows:
\begin{eqnarray}
{\cal L}_{ct}&=&Y_A\langle V_{\mu\nu}V^{\mu\nu}\chi_+\rangle -\frac{1}{2}Y_B\langle \nabla^{\lambda}V_{\lambda\mu}\nabla_{\nu}V^{\nu\mu}\chi_+\rangle \nonumber\\
&+&\frac{Y_{C_1}}{2}\langle \nabla^2V^{\mu\nu}\{\chi_+,\{\nabla_\nu,\nabla^\sigma\}V_{\mu\sigma}\}\rangle \nonumber\\
&+&\frac{Y_{C_2}}{4}\langle \{\nabla_\nu,\nabla_\alpha\}V^{\mu\nu}\{\chi_+,\{\nabla^\sigma,\nabla^\alpha\}V_{\mu\sigma}\}\rangle \nonumber\\
&+&\frac{Y_{C_3}}{4}\langle \{\nabla^\sigma,\nabla^\alpha\}V^{\mu\nu}\{\chi_+,\{\nabla_\nu,\nabla_\alpha\}V_{\mu\sigma}\}\rangle \nonumber\\
&+&\frac{Z_A F_V}{2\sqrt{2}}\langle V_{\mu\nu}f_{+}^{\mu\nu}\rangle +\frac{Z_B
F_V}{2\sqrt{2}}\langle V_{\mu\nu}\nabla^{2}f_{+}^{\mu\nu}\rangle \nonumber\\
&+&\frac{Z_C
F_V}{2\sqrt{2}}\langle V_{\mu\nu}\nabla^{4}f_{+}^{\mu\nu}\rangle +\frac{Z_D
F_V}{2\sqrt{2}}\langle V_{\mu\nu}\nabla^{6}f_{+}^{\mu\nu}\rangle \,.
\end{eqnarray}
We  adopt the $\overline{\rm MS}-1$ subtraction scheme and
absorb the divergent pieces proportional to $\lambda_\infty$ by the bare couplings in the counterterms.  Consequently, the remanent finite pieces of counterterms can be written as:
\bea
\Pi_{\rho\omega}^{ct}=X_W^r\, p^6+X_Z^r\,p^4+X_R^r\, p^2 \ ,
\eea
with
\bea
X_W^r&\equiv& \frac{8\pi\alpha F_\rho F_\omega}{3}(Z_D^r +Z_B^r
Z_C^r)\ ,\nonumber\\
 X_Z^r&\equiv& \frac{4\pi\alpha F_\rho F_\omega}{3}(2 Z_C^r
+{Z_B^r}^2)\nonumber\\
&&+16M_\rho(m_u-m_d)(Y_{C_1}^r+Y_{C_2}^r+Y_{C_3}^r)\ ,\nonumber\\
X_R^r&\equiv&\frac{8\pi\alpha F_\rho
F_\omega}{3}Z_B^r-4M_\rho(m_u-m_d)Y_B^r\ . \eea

In summary, at the NLO in $1/N_C$, the
UV-renormalized mixing amplitude reads
\begin{eqnarray} \label{eq.MixingAmplitude}
{\Pi}^r_{\rho\omega}(p^2)&=& S_{\rho\omega}^{(a)}\sqrt{Z_\rho}+\bar{S}_{\rho\omega}^{(c)}+\bar{S}_{\rho\omega}^{(d)}+\bar{S}_{\rho\omega}^{(e)}+\overline{E}_{\rho\omega
}^{\pi\gamma}(p^2)+X_W^r p^6+ X_Z^rp^4+X_R^rp^2\,,
\end{eqnarray}
where a bar indicates that the divergences are subtracted.

As discussed in Ref.~\cite{Connell97}, the mixing amplitude should vanish as $p^2\to 0$. Thus, the final expression of the renormalized mixing amplitude is obtained as follows:
\bea\label{eq.Pirhoomega}
{\Pi}_{\rho\omega}(p^2)={\Pi}^r_{\rho\omega}(p^2)-{\Pi}^r_{\rho\omega}(0)\ ,
\eea
where an additional finite shift is imposed to guarantee that the constraint ${\Pi}_{\rho\omega}(0)=0$ is satisfied.
Note that due to the finite shift performed in Eq.~\eqref{eq.Pirhoomega}, our numerical calculation is actually independent of the coupling $X_Y^r$.
In our numerical computation, the scale $\mu$ will be set to
$M_\rho$ and we use $(m_u-m_d)=-2.49$ MeV provided by PDG~\cite{ParticleDataGroup:2020ssz}.

\end{document}